\documentclass[%
 aip,
 amsmath,amssymb,
reprint, onecolumn 
]{revtex4-1}

\usepackage{graphicx}
\usepackage{dcolumn}

\usepackage{bm}
\bmdefine{\ba}{a}
\bmdefine{\bb}{b}
\bmdefine{\bc}{c}
\bmdefine{\bd}{d}
\bmdefine{\be}{e}
\bmdefine{\bf}{f}
\bmdefine{\bg}{g}
\bmdefine{\bj}{j}
\bmdefine{\bk}{k}
\bmdefine{\bn}{n}
\bmdefine{\bp}{p}
\bmdefine{\bq}{q}
\bmdefine{\br}{r}
\bmdefine{\bx}{x}
\bmdefine{\by}{y}
\bmdefine{\bu}{u}
\bmdefine{\bv}{v}
\bmdefine{\bw}{w}
\bmdefine{\bz}{z}
\bmdefine{\bE}{E}
\bmdefine{\bB}{B}
\bmdefine{\bC}{C}
\bmdefine{\bD}{D}
\bmdefine{\bJ}{J}
\bmdefine{\bR}{R}
\bmdefine{\bS}{S}
\bmdefine{\bT}{T}

\usepackage[utf8]{inputenc}
\usepackage[T1]{fontenc}
\usepackage{mathptmx}
\usepackage{etoolbox}

\usepackage{breqn}

\usepackage{cleveref}
\crefformat{pluralequation}{#2eqs.~(#1)#3}
\Crefformat{pluralequation}{#2Eqs.~(#1)#3}

\makeatletter
\newcommand*{\defeq}{\mathrel{\rlap{%
					 \raisebox{0.3ex}{$\m@th\cdot$}}%
					 \raisebox{-0.3ex}{$\m@th\cdot$}}%
					 =}
\makeatother

\def\foo#1{\xfoo#1\relax^\relax\valign}
\def\xfoo#1^#2\relax#3\valign{%
\mathbf{#1}\ifx\valign#2\valign\else^{\mathbf{#2}}\fi}

\makeatletter
\def\@email#1#2{%
 \endgroup
 \patchcmd{\titleblock@produce}
  {\frontmatter@RRAPformat}
  {\frontmatter@RRAPformat{\produce@RRAP{*#1\href{mailto:#2}{#2}}}\frontmatter@RRAPformat}
  {}{}
}%
\makeatother
\begin{document}


\title[A new class of special functions arising in plasma linear susceptibility tensor calculations]{A new class of special functions arising in plasma linear susceptibility tensor calculations}
\author{R. Ricci}
\email{roberto.ricci@enea.it.}
\affiliation{
ENEA Nuclear Department \\Frascati Research Center Via E. Fermi 45 00044 Frascati RM Italy
}%

\date{\today}

\begin{abstract}
We investigate some fundamental properties of a peculiar class of special functions strictly related to Bessel, Anger and Weber functions, whose introduction was originally motivated by linear susceptibility tensor calculations in a hot, magnetised plasma. We show that these functions are solutions of an inhomogeneous Bessel ODE, with specified initial conditions and a distinct right-hand-side term fulfilling the Nielsen's requirement. Beside deriving recurrence relations and an alternative representation involving incomplete Anger-Weber functions, we show that these functions admit a simple series expansion in terms of Bessel functions of integer order, obtained by resorting to the Jacobi-Anger formula. In plasma applications this eventually leads to expressions involving infinite sums of products of Bessel functions, not particularly apt to numerical evaluation ought to their slow convergence rate when the particle's gyro-radius is larger than the wavelength. By exploiting the previously determined recurrence properties of the new class of functions we present a particularly simple derivation of the linear susceptibility tensor that enables to avoid this inconvenience. 
\end{abstract}

\maketitle

\section{\label{sec:introduction}Introduction
}

Some years ago Qin, Philips and Davidson \cite{hQ07} proposed a method to avoid the infinite sums of Bessel functions that usually plague kinetic model  calculations of the linear equivalent susceptibility tensor of a hot magnetised plasma. Their proposal promised to be particularly advantageous in view of numerical implementations, especially in the regime of large particle gyro-radius. In fact, in that regime the rate of convergence of the  arising series of Bessel functions is very slow, and this forces to include a huge number of terms in numerical computations.

The method relies on the introduction of a function, defined as a particular integral of Bessel type, whose remarkable properties enable to simplify calculations by avoiding the recourse to the Jacobi-Anger formula. Indeed, it is exactly the use of this formula in the traditional approach \cite{thS92} that eventually causes the arising of series of Bessel functions in final results. It is important to remark that, as shown by Swanson \cite{dgS03}, all the resultant series can be ultimately summed up explicitly by using a sum rule found by Newberger in 1982 \cite{bsN82}, but the overall procedure results to be quite cumbersome and, given the algebraic complexity of the calculations, prone to errors.

In the present article we clarify the nature of the function defined by Qin et al., showing that it is the solution of the inhomogeneous Bessel function with a particular right-hand-side term and well-specified initial conditions. We manage to give it an explicit analytic expression in terms of known special functions and investigate some of its more relevant properties, including recurrence relations and series expansion in terms of Bessel functions.

We also provide an application of these new theoretical findings to plasma physics, presenting a particularly simple new method for the calculation of the linear susceptibility tensor that systematises and generalises the results obtained by Qin \emph{et al.}. A thorough discussion of the connections of our findings with the Newberger's sum rule is given in appendix.

\section{Definitions and preliminaries}\label{sec:preliminaries}

Following Qin \emph{et al.} \cite{hQ07}, we consider the function defined by the improper integral:

\begin{dmath}\label{eq:G}
	G_\mu(z, \psi) \defeq i\omega_c\int_0^{\infty}\mathrm{d}\tau\,\mathrm{e}^{-i z \sin( \omega_c\tau + \psi) - i  \mu\omega_c\tau}
\end{dmath},
where $z$ and $\psi$ are real variables, $\mu$ is in general a complex non-integer constant parameter and $\omega_c$ is a non-zero real  constant\footnote{In the context of plasma kinetic theory, where such function naturally pops up in the calculation of the susceptibility tensor,  $z \defeq k_\perp v_\perp / \omega_c$, with $\omega_c \defeq qB_0/m$ the signed cyclotron frequency of a particle of mass $m$ and charge $q$ subject to an external constant magnetic field of intensity $B_0$ and $\mu \defeq (k_\parallel v_\parallel - \Omega)/\omega_c$, where $\Omega \defeq \omega + i \gamma$ is the complex frequency, comprising the real frequency $\omega$ and the dumping (or growing, depending on the sign) factor $\gamma$.}.
By exploiting the evident periodicity of $G_\mu(z,\psi)$ in the variable $\psi$,  with period equal to $2\pi$,

\begin{dmath}
G_\mu(z, \psi +2\pi) \equiv G_\mu(z, \psi)
\end{dmath}, 
\cref{eq:G} can also be conveniently reformulated in terms of a proper definite integral. Namely, since

\begin{dmath}
	G_\mu(z, \psi) = i\int_0^{\infty}\mathrm{d}u\,\mathrm{e}^{-i z \sin( u + \psi) - i  \mu u} = 
	G_\mu(z, \psi + 2\pi) = 
	i\mathrm{e}^{i 2\pi \mu}\int_0^{\infty}\mathrm{d}u \,\mathrm{e}^{-i z \sin( u + \psi + 2\pi) - i  \mu (u+2\pi)} = 
	i\mathrm{e}^{i 2\pi \mu} \int_{2\pi}^{\infty}\mathrm{d}\lambda\,\mathrm{e}^{-i z \sin( \lambda + \psi) - i  \mu \lambda} = 
	\mathrm{e}^{i2\pi \mu}\left( G_\mu(z, \psi) - i\int_0^{2\pi}\mathrm{d}\lambda\,\mathrm{e}^{-i z \sin( \lambda + \psi) - i  \mu \lambda}\right)
\end{dmath},
we have the alternative formulation, which we assume as the primary definition of $G_\mu(z, \psi)$:

\begin{dmath}\label{eq:G-definite}
	G_\mu(z, \psi) =  c_\mu\int_0^{2\pi}\frac{\mathrm{d}\lambda}{2\pi}\,\mathrm{e}^{-i z \sin( \lambda + \psi) - i  \mu \lambda} 
\end{dmath},
where
\begin{dmath}\label{eq:c_mu}
	c_\mu  \defeq \frac{\pi\mathrm{e}^{i\pi \mu}}{\sin\pi \mu} 
\end{dmath}.
We note that:

\begin{dgroup}		
	\begin{dmath}\label{eq:G-z-derivative}
		\partial_z G_\mu(z, \psi) = \omega_c\int_0^{\infty}\mathrm{d}\tau\,\mathrm{e}^{-i z \sin( \omega_c\tau + \psi) - i  \mu\omega_c\tau} \sin(\omega_c\tau + \psi) =
		-ic_\mu\int_0^{2\pi}\frac{\mathrm{d}\lambda}{2\pi}\,\mathrm{e}^{-i z \sin( \lambda + \psi) - i  \mu \lambda}\sin(\lambda + \psi)
	\end{dmath},
	\begin{dmath}\label{eq:G-psi-derivative}
		\frac{i}{z}\partial_\psi G_\mu(z, \psi) = i\omega_c\int_0^{\infty}\mathrm{d}\tau\,\mathrm{e}^{-i z \sin( \omega_c\tau + \psi) - i  \mu\omega_c\tau} \cos(\omega_c\tau + \psi) =
		c_\mu\int_0^{2\pi}\frac{\mathrm{d}\lambda}{2\pi}\,\mathrm{e}^{-i z \sin( \lambda + \psi) - i  \mu \lambda}\cos(\lambda + \psi)
	\end{dmath}.
\end{dgroup}
By combining \cref{eq:G-z-derivative} and \cref{eq:G-psi-derivative} we obtain:

\begin{dmath}\label{eq:G-derivatives}
	\left(\partial_{z} \pm \frac{i}{z}\partial_{\psi}\right) G_\mu(z, \psi) = \pm i\omega_c\int_0^{\infty}\mathrm{d}\tau\,\mathrm{e}^{-i z \sin( \omega_c\tau + \psi) - i  \mu\omega_c\tau} \,\mathrm{e}^{\mp i(\omega_c\tau + \psi)} = 
	\pm c_\mu\int_0^{2\pi}\frac{\mathrm{d}\lambda}{2\pi}\,\mathrm{e}^{-i z \sin( \lambda + \psi) - i  \mu \lambda}\,\mathrm{e}^{\mp i(\lambda + \psi)} =
	\pm \mathrm{e}^{\mp i\psi}c_\mu \int_0^{2\pi}\frac{\mathrm{d}\lambda}{2\pi}\,\mathrm{e}^{-i z \sin( \lambda + \psi) - i(\mu \pm 1) \lambda}
\end{dmath}.
In summary, considering that $c_\mu$ is a periodic function of $\mu$ of period $1$ (i.e. $c_{\mu \pm 1} =c_\mu$) the functions $G_\mu(z, \psi)$ satisfy the following differential-recursive relations:
\begin{dgroup}\label[pluralequation]{eq:G-derivatives+-}
	\begin{dmath}\label{eq:G-deriv+}
		\left(\partial_{z} + \frac{i}{z}\partial_{\psi}\right) G_\mu(z, \psi) = \mathrm{e}^{-i\psi}G_{\mu + 1}(z, \psi)
	\end{dmath},
	\begin{dmath}\label{eq:G-deriv-}
		\left(\partial_{z} - \frac{i}{z}\partial_{\psi}\right) G_\mu(z, \psi) = -\mathrm{e}^{i\psi}G_{\mu - 1}(z, \psi)
	\end{dmath}.
\end{dgroup}

\section{Properties of $G_\mu(z, \psi)$} 
In the next subsections we discuss some fundamental properties of the just introduced class of functions. 

\subsection{Expansion in terms of cylindrical Bessel functions}

If we expand the complex exponentials using the Jacobi-Anger identity\footnote{Unless explicitly stated otherwise, we adopt throughout the shortcut $\sum_n$ for $\sum_{n=-\infty}^\infty$.}:

\begin{dmath}\label{eq:Jacobi-Anger}
	\mathrm{e}^{\pm i z \sin \psi} = \sum_n J_n(z)\mathrm{e}^{\pm in\psi}
\end{dmath},
where $J_n(z)$ denotes the cylindrical Bessel function of integer order $n$, the integrals in \cref{eq:G-definite} and \cref{eq:G-derivatives} can be carried out explicitly. It is immediate to verify that

\begin{dgroup}\label[pluralequation]{eq:G-B}
	\begin{dmath}\label{eq:G-bessel}
		G_\mu(z, \psi) = \sum_n \frac{J_n(z)}{n + \mu}\mathrm{e}^{-in\psi}
	\end{dmath}
	\begin{dmath}\label{eq:G-derivatives-bessel}
		\left(\partial_{z}\right \pm \frac{i}{z}\partial_{\psi}) G_\mu(z, \psi) = \pm\sum_n \frac{J_{n\mp 1}(z)}{n + \mu}\mathrm{e}^{-in\psi} =
		{\pm\mathrm{e}^{\mp i\psi}\sum_n \frac{J_{n}(z)}{n +(\mu \pm 1)}\mathrm{e}^{-in\psi} \equiv }\pm \mathrm{e}^{\mp i\psi}G_{\mu \pm 1}(z, \psi)
	\end{dmath},
\end{dgroup}
consistently with \cref{eq:G-derivatives+-}.
\Cref{eq:G-derivatives-bessel} can also be obtained directly by \cref{eq:G-bessel} by deriving the series term by term. We note \emph{en passant} that \cref{eq:G-bessel} can also be interpreted as the Fourier series expansion in $\psi$ of the periodic function $G_\mu(z, \psi)$, with Fourier coefficients:

		\begin{dmath}
			\hat{G}_{\mu,n}(z) \defeq \frac{J_n(z)}{n+\mu}
		\end{dmath}.

Both \cref{eq:G-bessel} and \cref{eq:G-derivatives-bessel} involve the following series of Bessel functions

\begin{dmath} \label{eq:S}
	\sum_n \frac{J_n(z)}{n+\mu}\mathrm{e}^{-in\psi} 
\end{dmath},
which is a special case of the most general series \cite{gD89}

\begin{dmath} \label{eq:S-general}
	S^{(m)}_{\mu}(z, t) \defeq \sum_n \frac{J^m_n(z)}{n+\mu}t^n 
\end{dmath}
defined for non-integer $\mu$, positive integer $m$ and complex $t$, $0<|t|<\infty$. Namely:

\begin{dmath}
	G_\mu(z, \psi) \equiv S^{(1)}_{\mu}(z,\mathrm{e}^{-i\psi})
\end{dmath}.

When $\psi=0$ (i.e. $t=1$), the resulting series $G_\mu(z, 0) \equiv S^{(1)}_{\mu}(z, 1)$ can be summed explicitly in terms of the Anger function \cite{gD89}, defined by:

\begin{dmath}
	\foo{J}_{\mu}(z) \defeq \int_{0}^{\pi} \frac{\mathrm{d}\lambda}{\pi}\cos(\mu\lambda - z\sin\lambda)
\end{dmath}.
It holds that:

\begin{dmath} \label{eq:S-Anger}
	G_\mu(z, 0) = {\sum_n \frac{J_n(z)}{n+\mu} = 
		\frac{\pi}{\sin\pi \mu}\foo{J}_{\mu}(z) }
\end{dmath}.
A derivation of \cref{eq:S-Anger}, as well an independent verification obtained directly from the definition \cref{eq:G-definite}, can be found in \cref{sec:S-Anger}.

\subsection{Recurrence relations}

Recurrence relations for the functions $G_{\mu}(z,\psi)$ can be derived starting from the corresponding well-known relations valid for Bessel functions of integer order:

\begin{dgroup}\label[pluralequation]{eq:Bessel-recurrence-rels}
	\begin{dmath}\label{eq:Bessel-recurrence-rels1}
		J_{n-1}(z)-J_{n+1}(z) - 2\partial_{z} J_n(z) = 0
	\end{dmath},
	\begin{dmath}\label{eq:Bessel-recurrence-rels1}
		J_{n-1}(z)+J_{n+1}(z) - \frac{2n}{z} J_n(z) = 0
	\end{dmath}.
\end{dgroup}
If we multiply by $\mathrm{e}^{-in\psi}/(n + \mu)$ and sum over $n$ we immediately  get:

\begin{dgroup}\label[pluralequation]{eq:G-recurrence}
	\begin{dmath}\label{eq:G-recurrence1}
		\mathrm{e}^{i\psi} G_{\mu - 1}(z, \psi) - \mathrm{e}^{-i\psi} G_{\mu + 1}(z, \psi) + 2\partial_z G_{\mu}(z, \psi) = 0
	\end{dmath},
		\begin{dmath}\label{eq:G-recurrence2}
		\mathrm{e}^{i\psi} G_{\mu - 1}(z, \psi) + \mathrm{e}^{-i\psi} G_{\mu + 1}(z, \psi)  + \frac{2\mu}{z}G_{\mu}(z, \psi) = \frac{2}{z} \mathrm{e}^{-iz\sin\psi}
	\end{dmath}.
\end{dgroup}
Once expressed in terms of $S^{(1)}_{\mu}(z, t)$, these relations read:

		\begin{dgroup}
			\begin{dmath}
				\frac{1}{t} S^{(1)}_{\mu - 1}(z, t) - t S^{(1)}_{\mu + 1}(z, t) + 2\partial_z S^{(1)}_{\mu}(z, t) = 0
			\end{dmath},
			\begin{dmath}
				\frac{1}{t} S^{(1)}_{\mu - 1}(z, t) + t S^{(1)}_{\mu + 1}(z, t) +\frac{2\mu}{z}S^{(1)}_{\mu}(z, t) = \frac{2}{z} \mathrm{e}^{\frac{z}{2}(t-\frac{1}{t})}
			\end{dmath}.
		\end{dgroup}
		
By setting $\psi=0$ and noticing that 

\begin{dmath}
	G_{\mu \pm 1}(z, 0) = {\frac{\pi}{\sin(\mu \pm 1)\pi}\foo{J}_{\mu \pm 1}(z) = -\frac{\pi}{\sin\mu\pi}\foo{J}_{\mu \pm 1}(z)}
\end{dmath},
after simplifying the common factor $\frac{\pi}{\sin\mu\pi}$ \cref{eq:G-recurrence} reduce as expected to the recurrence relations for the Anger functions, namely:

\begin{dgroup}
	\begin{dmath}
		 \foo{J}_{\mu - 1}(z) - \foo{J}_{\mu + 1}(z) - 2\partial_z \foo{J}_{\mu}(z) = 0
	\end{dmath}
	\begin{dmath}
			\foo{J}_{\mu - 1}(z) + \foo{J}_{\mu + 1}(z) - \frac{2\mu}{z}\foo{J}_{\mu}(z) = -\frac{2}{\pi z} \sin\mu\pi
	\end{dmath}.
\end{dgroup}
In view of further developments, it is interesting to note that, if we tentatively define:

\begin{dmath}\label{eq:G-tilde}
	\tilde{G}_{\mu}(z, \psi) \defeq \frac{\sin\mu\pi}{\pi}\mathrm{e}^{-i\mu\psi} G_{\mu}(z, \psi)
\end{dmath},
the new rescaled functions satisfy the following more canonical recurrence relations:

\begin{dgroup}\label[pluralequation]{eq:G-tilde-recurrence}
	\begin{dmath}\label{eq:G-tilde-recurrence1}
		\tilde{G}_{\mu - 1}(z, \psi) - \tilde{G}_{\mu + 1} (z, \psi) - 2\partial_z \tilde{G}_{\mu}(z, \psi) = 0
	\end{dmath},
		\begin{dmath}\label{eq:G-tilde-recurrence2}
		\tilde{G}_{\mu - 1}(z, \psi) + \tilde{G}_{\mu + 1}(z, \psi) -\frac{2\mu}{z}\tilde{G}_{\mu}(z, \psi) =  \frac{2}{z} \tilde{g}_\mu(z, \psi)
	\end{dmath},
\end{dgroup}
where

\begin{dmath}\label{eq:g-tilde}
	\tilde{g}_\mu(z, \psi) \defeq -\frac{\sin\mu\pi}{\pi}\mathrm{e}^{-iz\sin\psi - i\mu\psi}
\end{dmath}.
It can be immediately verified hat the functions $\tilde{g}_\mu(z, \psi)$, in turn, fulfil the equation:

\begin{dmath}\label{eq:g-tilde-Nielsen}
		\tilde{g}_{\mu - 1}(z, \psi) - \tilde{g}_{\mu + 1}(z, \psi) - 2\partial_z \tilde{g}_{\mu}(z, \psi) = 0
\end{dmath}.

\subsection{Connection with Nielsen's problem} \label{sec:Nielsen}

\noindent In the classic literature on Bessel functions (see e.g. \cite{mmA71} \cite{nW66}) the problem is widely discussed whether necessary and sufficient conditions exist that enable to find solutions $E_\mu(z)$ to the following so-called Nielsen's functional equations, which represent a generalization of the recurrence relations \cref{eq:Bessel-recurrence-rels} fulfilled by cylindrical functions:

\begin{dgroup}\label[pluralequation]{eq:Nielsen-feq}
	\begin{dmath}\label{eq:Nielsen-feq1}
		E_{\mu-1}(z) - E_{\mu+1}(z) - 2\partial_{z} E_\mu(z) = \frac{2}{z} f_\mu(z)
	\end{dmath},
	\begin{dmath}\label{eq:Nielsen-feq2}
		E_{\mu-1}(z) + E_{\mu+1}(z) - \frac{2\mu}{z} E_\mu(z) = \frac{2}{z} g_\mu(z)
	\end{dmath}.
\end{dgroup}
It is possible to demonstrate that a solution exists if and only if the two otherwise arbitrary functions $g_\mu(z)$ and $f_\mu(z)$ fulfil the so-called Nielsen's condition:

\begin{dmath}\label{eq:Nielsen-condition}
		g_{\mu - 1}(z, \psi) - g_{\mu + 1}(z, \psi) - 2\partial_{z} g_{\mu}(z, \psi) = f_{\mu - 1}(z, \psi) - f_{\mu + 1}(z, \psi) - 2\partial_{z} f_{\mu}(z, \psi)
\end{dmath}.
Specifically, if \cref{eq:Nielsen-condition} is satisfied, then the system of \cref{eq:Nielsen-feq} is equivalent to an inhomogeneous Bessel differential equation in which the right side can be expressed only in terms of the functions $g_\mu(z)$ and $f_\mu(z)$. Namely, adopting the notation $\mathcal{D}_{B_{z,\mu}}$ for the Bessel differential operator:

\begin{dmath}\label{eq:Bessel-diff-op}
	 \mathcal{D}_{B_{z,\mu}} \defeq z^2\partial_{z}^2 + z\partial_{z} + z^2 - \mu^2
\end{dmath},
\cref{eq:Nielsen-feq} and \cref{eq:Nielsen-condition} entail:

\begin{dmath}\label{eq:Bessel-inhom}
	\mathcal{D}_{B_{z,\mu}}E_{\mu}(z, \psi) = \ell_\mu(z, \psi),
\end{dmath}
where:

\begin{dmath}
	\ell_\mu(z) \defeq -z [f_{\mu - 1}(z) - g_{\mu - 1}(z)] - (z\partial_z - \mu)[f_{\mu}(z) + g_{\mu}(z)].
\end{dmath}
 If we take the Bessel functions of the first kind $J_\mu(z)$ and $J_{-\mu}(z)$ as a pair of independent solutions of the correspondent homogeneous problem, the general solution $E_\mu(z, \psi)$ of \cref{eq:Bessel-inhom}, found by the method of variation of parameters, is given by: 

\begin{dmath} 
	E_\mu(z) = J_{\mu}(z) \left[A_\mu - \int_a^z d\zeta\frac{J_{-\mu}(\zeta)\ell_\mu(\zeta)}{\zeta^2 W_{J_\mu, J_{-\mu}}(\zeta)}\right] + \\
	J_{-\mu}(z) \left[B_\mu + \int_b^z d\zeta\frac{J_{\mu}(\zeta)\ell_\mu(\zeta)}{\zeta^2 W_{J_\mu, J_{-\mu}}(\zeta)}\right]
\end{dmath},
where

\begin{dmath}\label{eq:wronskian}
	W_{J_\mu, J_{-\mu}}(z) \defeq {J_\mu(z) J'_{-\mu}(z) - J'_\mu(z) J_{-\mu}(z) = -\frac{2}{\pi z} \sin{\pi\mu}}
\end{dmath}
is the Wronskian of the functions $J_\mu(z)$ and $J_{-\mu}(z)$ and $A_\mu$, $B_\mu$, $a$, $b$ are arbitrary constants, not depending on $z$.

Alternatively, by taking $J_\mu(z)$ and the Neumann function $Y_{\mu}(z)$ as an independent pair, the general solution $E_\mu(z, \psi)$ is given by: 

\begin{dmath}
	E_\mu(z) = J_{\mu}(z) \left[\bar{A}_\mu - \int_{\bar{a}}^z d\zeta\frac{Y_{\mu}(\zeta)\ell_\mu(\zeta)}{\zeta^2 W_{J_\mu, Y_{\mu}}(\zeta)}\right] + \\
		Y_{\mu}(z) \left[\bar{B}_\mu + \int_{\bar{b}}^z d\zeta\frac{J_{\mu}(\zeta)\ell_\mu(\zeta)}{\zeta^2 W_{J_\mu, Y_{\mu}}(\zeta)}\right]
\end{dmath},
w here

\begin{dmath}
W_{J_\mu, Y_{\mu}}(z) = \frac{2}{\pi z}
\end{dmath}
is the Wronskian of the functions $J_\mu(z)$ and $Y_{\mu}(z)$. 

It is immediate to realise that, apart from the harmless dependence of $\tilde{G}_{\mu}$ and $\tilde{g}_{\mu}$ on the additional parameter $\psi$, \cref{eq:G-tilde-recurrence} and \cref{eq:g-tilde-Nielsen} respectively correspond to \cref{eq:Nielsen-feq} and \cref{eq:Nielsen-condition} in the special case $f_\mu(z) \equiv 0$. Note that in such a case the Nielsen's condition becomes:
\begin{dmath}\label{eq:Nielsen-condition-f=0}
		g_{\mu - 1}(z, \psi) - g_{\mu + 1}(z, \psi) - 2\partial_{z} g_{\mu}(z, \psi) = 0
\end{dmath}
and the definition of $\ell_\mu(z)$ simplifies to:

\begin{dmath}\label{eq:ell-f=0}
	\ell_\mu(z) \defeq z g_{\mu - 1}(z) - (z\partial_z - \mu) g_{\mu}(z).
\end{dmath}
Indeed, by summing and subtracting \cref{eq:G-tilde-recurrence} we obtain:

\begin{dgroup}
	\begin{dmath}
		(z\partial_z - \mu)\tilde{G}_{\mu}(z, \psi) = -z \tilde{G}_{\mu + 1}(z, \psi) + \tilde{g}_{\mu}(z, \psi)
	\end{dmath},
	\begin{dmath}
		(z\partial_z + \mu)\tilde{G}_{\mu}(z, \psi) = z \tilde{G}_{\mu - 1}(z, \psi) - \tilde{g}_{\mu}(z, \psi)
	\end{dmath},
\end{dgroup}
and if we apply the operator ($z\partial_z + \mu)$ to the first equation and use the second to eliminate $\tilde{G}_{\mu + 1}(z, \psi)$ we eventually get, as preannounced:

\begin{dmath}\label{eq:Gessel-tilde}
	\mathcal{D}_{B_{z,\mu}}\tilde{G}_{\mu}(z, \psi) = {z \tilde{g}_{\mu - 1}(z, \psi) - (z\partial_z - \mu)\tilde{g}_{\mu}(z, \psi) \defeq \tilde{\ell}_\mu(z, \psi)}.
\end{dmath}
From \cref{eq:g-tilde} we can derive in the present case the specific expression:

\begin{dmath}\label{eq:Gessel-tilde-rhs}
	\tilde{\ell}_\mu(z, \psi) = 
		\frac{\sin\mu\pi}{\pi}\mathrm{e}^{-iz\sin\psi -i\mu\psi}(z\cos\psi - \mu)
\end{dmath}.
Taking into account the definition of $\tilde{G}_{\mu}(z, \psi)$, \cref{eq:G-tilde}, this obviously entails that also the original function $G_{\mu}(z, \psi)$ is solution of a particular inhomogeneous Bessel ODE, namely:
\begin{dmath}\label{eq:Gessel}
	\mathcal{D}_{B_{z,\mu}}G_{\mu}(z, \psi) = \ell_\mu(z, \psi)
\end{dmath},
with

\begin{dmath}\label{eq:Gessel-rhs}
	\ell_\mu(z, \psi) \defeq 
		\mathrm{e}^{-iz\sin\psi}(z\cos\psi - \mu)
\end{dmath}.

\subsection{Governing differential equations} \label{sec:diff-eq}

In the previous subsection, by exploiting the strict analogy of \cref{eq:G-tilde-recurrence} and \cref{eq:g-tilde-Nielsen} with the classical Nielsen's problem, we have established that $G_{\mu}(z, \psi)$ is a specific solution of the inhomogeneous Bessel equation whose right-hand side is given by \cref{eq:Gessel-rhs}. 

\noindent It is interesting to notice that this result could have also been obtained by directly applying the Bessel operator to \cref{eq:G-definite}. Namely, on the one hand we have:

\begin{dmath}\label{eq:first-term}
	\mathcal{D}_{B_{z,\mu}}G_\mu(z, \psi) = 
	c_\mu\int_0^{2\pi}\frac{\mathrm{d}\lambda}{2\pi}\mathrm{e}^{-i z \sin(\lambda + \psi) - i  \mu \lambda}[z^2\cos^2(\lambda + \psi) - \mu^2 - iz\sin(\lambda + \psi)]
\end{dmath}.
On the other hand, if we apply the auxiliary differential operator $\mathcal{D}_{A_{\lambda,\mu}} \defeq -(\partial^2_\lambda + 2i\mu\partial_\lambda)$ on the exponential function appearing in \cref{eq:first-term} we obtain: 

\begin{dmath}\label{eq:second-term}
	\mathcal{D}_{A_{\lambda,\mu}}\mathrm{e}^{-i z \sin( \lambda + \psi) - i  \mu \lambda} = 
	\mathrm{e}^{-i z \sin( \lambda + \psi) - i  \mu \lambda} [z^2 \cos^2(\lambda + \psi) - \mu^2 - iz\sin(\lambda + \psi)]
\end{dmath}.
Thus, by direct comparison of \cref{eq:first-term} and \cref{eq:second-term}, we obtain

\begin{dmath}
	\mathcal{D}_{B_{z,\mu}}G_\mu(z, \psi) = 
		c_\mu\int_0^{2\pi}\frac{\mathrm{d}\lambda}{2\pi}\mathcal{D}_{A_{\lambda,\mu}}\mathrm{e}^{-i z \sin( \lambda + \psi) - i  \mu \lambda} = 
	\left[-\frac{c_\mu}{2\pi}(\partial_\lambda + 2i\mu)\mathrm{e}^{-i z \sin( \lambda + \psi) - i  \mu \lambda}\right]\bigg\vert_0^{2\pi} =
			\left\{\frac{c_\mu}{2\pi}\mathrm{e}^{-i z \sin( \lambda + \psi) - i  \mu \lambda} [iz\cos(\lambda + \psi) - i\mu]\right\}\bigg\vert_0^{2\pi} =
	\mathrm{e}^{-i z \sin\psi}(z\cos\psi - \mu)
\end{dmath},
as expected.

\noindent In particular, using the series representation \cref{eq:G-bessel}, it is easy to demonstrate that $G_\mu(z, \psi)$ is the peculiar solution subject to the initial conditions:

\begin{dgroup}\label[pluralequation]{eq:ic-z}
	\begin{dmath}
		G_{\mu}(0, \psi)  = {\frac{1}{\mu} \equiv -\frac{1}{\mu^2}}\ell_\mu(0, \psi)
	\end{dmath},
		\begin{dmath}
		\partial_{z}G_{\mu}(0,\psi) = {\frac{\cos\psi + i\mu\sin\psi}{1 - \mu^2} \equiv -\frac{1}{\mu^2 - 1}\partial_z\ell_\mu(0, \psi)}
	\end{dmath}.
\end{dgroup}

On the other hand, if we start from \cref{eq:G-derivatives+-}, rearranged as:

\begin{dgroup}
	\begin{dmath}
		\left(z\partial_z + i\partial_\psi\right) G_\mu(z, \psi) = \mathrm{e}^{-i\psi}zG_{\mu + 1}(z, \psi)
	\end{dmath},
	\begin{dmath}
		\left(z\partial_z - i\partial_\psi\right) G_\mu(z, \psi) = -\mathrm{e}^{i\psi}zG_{\mu - 1}(z, \psi)
	\end{dmath},
\end{dgroup}
and apply the differential operator $(z\partial_{z} - i\partial_{\psi})$ to the first equation, after some simple algebra we eventually get the following second-order PDE for $G_\mu(z, \psi)$:

\begin{dmath}\label{eq:G-diff-eq}
	 (z^2\partial_{z}^2 + z\partial_{z} + z^2) G_{\mu}(z,\psi) = 
		-\partial_{\psi}^2 G_{\mu}(z, \psi)
\end{dmath},
which can be solved by separation of variables. Indeed, it is not difficult to verify that, for each integer $n$ the particular product of $z$ and $\psi$ functions $J_n(z)\mathrm{e}^{-in\psi}$ is a solution of \cref{eq:G-diff-eq}. By linearity, this entails, as expected, that the series \cref{eq:G-bessel} is also a solution, specifically the solution satisfying the initial conditions \cref{eq:ic-z} as well as the additional conditions:

\begin{dgroup}\label[pluralequation]{eq:ic-psi}
	\begin{dmath} \label{eq:psi=0}
		G_{\mu}(z, 0) =	\frac{\pi}{\sin\pi \mu}\foo{J}_{\mu}(z)
	\end{dmath},
	\begin{dmath}
		\partial_{\psi}G_{\mu}(z, 0) = -i + i \frac{\pi\mu}{\sin\pi \mu}\foo{J}_{\mu}(z)
	\end{dmath}.
\end{dgroup}

In order to derive an alternative integral representation for $G_\mu(z, \psi)$, it is useful to follow a different approach. If we add the term $-\mu^2 G_\mu(z, \psi)$ to both sides of \cref{eq:G-diff-eq}, we obviously get:

\begin{dmath}\label{eq:GG}
	 \mathcal{D}_{B_{z,\mu}}G_\mu(z, \psi) = 
		-(\partial_{\psi}^2 + \mu^2)G_{\mu}(z, \psi)
\end{dmath}.
So, recalling \cref{eq:Gessel} and by virtue of \cref{eq:GG}, we have established that the function  $G_\mu(z, \psi)$ is also solution of the much simpler second-order inhomogeneous ODE in $\psi$:

\begin{dmath}\label{eq:G-ODE}
	(\partial^2_\psi + \mu^2)G_\mu(z, \psi) = -\ell_\mu(z, \psi)
\end{dmath},
subject to the initial conditions \cref{eq:ic-psi}.
The solution of \cref{eq:G-ODE} fulfilling all the initial conditions can be determined uniquely, e.g. by using the Green's function method to find a particular integral of the inhomogeneous problem. The final result reads:

\begin{dmath}
	G_\mu(z, \psi) = -i \frac{\sin\mu\psi}{\mu} + \frac{\pi}{\sin\pi \mu}\foo{J}_{\mu}(z) \mathrm{e}^{i \mu\psi} + \frac{1}{\mu}\int_0^\psi \mathrm{d}\lambda \sin(\mu(\psi - \lambda)) \mathrm{e}^{-i z \sin\lambda}(\mu - z \cos\lambda)
\end{dmath}.
This result is consistent with the solution found in the special case $\psi=0$ (cfr.  \cref{sec:S-Anger}) and satisfies all initial conditions in both variables, \cref{eq:ic-z} and \cref{eq:ic-psi}, as can be straightforwardly yet laboriously verified.
An alternative, simpler formulation can be derived by rewriting the factor $\sin(\mu(\psi - \lambda))$ in terms of complex exponentials. Namely:

\begin{dmath}\label{eq:G-ODE-simpler}
		G_\mu(z, \psi) = \mathrm{e}^{i \mu\psi}\left[\frac{\pi}{\sin\pi \mu}\foo{J}_{\mu}(z)  -2\pi i \int_0^{\psi}\frac{\mathrm{d}\lambda}{2\pi}  \mathrm{e}^{-i z \sin\lambda - i \mu\lambda}\right]
\end{dmath}.

We remark that, as an immediate consequence of the definition  \cref{eq:G-tilde}, we also have
\begin{dmath}\label{eq:G-tilde-ODE-simpler}
		\tilde{G}_\mu(z, \psi) = \foo{J}_{\mu}(z)  -2i\sin\mu\pi \int_0^{\psi}\frac{\mathrm{d}\lambda}{2\pi}  \mathrm{e}^{-i z \sin\lambda - i \mu\lambda}
\end{dmath}.
The $2\pi$-periodicity in $\psi$ of \cref{eq:G-ODE-simpler}, though not immediately apparent, can be easily checked by exploiting the identity (see \cref{sec:S-Anger}):
	\begin{dmath}
		\mathrm{e}^{i\mu\pi}\int_0^{2\pi} \frac{\mathrm{d}\lambda}{2\pi}  \mathrm{e}^{-i z \sin\lambda - i \mu\lambda} = \foo{J}_{\mu}(z)
	\end{dmath}.
	Note that $\tilde{G}_\mu(z, \psi)$ is not periodic in $\psi$, but rather satisfies the relation:
	\begin{dmath}
		\tilde{G}_\mu(z, \psi + 2\pi) =  \mathrm{e}^{-2\pi i\mu}\tilde{G}_\mu(z, \psi)
	\end{dmath}.

\section{Connections with the incomplete Anger-Weber functions}

A direct connection exists between the functions $G_\mu(z,\psi)$ investigated in this article and the class of incomplete Anger-Weber functions \cite{mmA71} \cite{dC19}. Incomplete Anger-Weber functions are characterized by the following Bessel-like integral representation, which can be assumed as a definition: 

\begin{dmath}
		\foo{A}_{\mu}(z, \psi) \defeq \int_0^{\psi} \frac{\mathrm{d}\lambda}{\pi}  \mathrm{e}^{i z \sin\lambda - i \mu\lambda} = 
		\int_0^{\psi} \frac{\mathrm{d}\lambda}{\pi}  \cos(\mu\lambda - z \sin\lambda) - i \int_0^{\psi} \frac{\mathrm{d}\lambda}{\pi}  \sin(\mu\lambda - z \sin\lambda) =
		\foo{J}_{\mu}(z, \psi) - i \foo{E}_{\mu}(z, \psi)
\end{dmath}.
$\foo{J}_{\mu}(z, \psi)$ and $\foo{E}_{\mu}(z, \psi)$ are respectively the incomplete Anger and Weber functions, which reduce to the corresponding "complete" versions when $\psi = \pi$:

\begin{dmath}
		\foo{A}_{\mu}(z, \pi)  =
		\foo{J}_{\mu}(z, \pi) - i \foo{E}_{\mu}(z, \pi) \equiv
		\foo{J}_{\mu}(z) - i \foo{E}_{\mu}(z)
\end{dmath},
thus justifying the use of the attribute "incomplete". Note that

\begin{dgroup}
	\begin{dmath}
		\foo{A}_{\mu}(-z, \psi) = \foo{J}_{-\mu}(z, \psi) + i \foo{E}_{-\mu}(z, \psi)
	\end{dmath},
	\begin{dmath}
		\foo{A}_{-\mu}(z, \psi) = \foo{J}_{-\mu}(z, \psi) - i \foo{E}_{-\mu}(z, \psi)
	\end{dmath}.

\end{dgroup}

\noindent The functions $\foo{A}_{\mu}(z, \psi)$ satisfy the following recurrence relations (cfr. \cref{eq:Nielsen-feq}):
\begin{dgroup}
	\begin{dmath}
		\foo{A}_{\mu - 1}(z, \psi) - \foo{A}_{\mu + 1}(z, \psi) - 2\partial_z \foo{A}_{\mu}(z, \psi) = 0
	\end{dmath}
	\begin{dmath}
			\foo{A}_{\mu - 1}(z, \psi) + \foo{A}_{\mu + 1}(z, \psi) - \frac{2\mu}{z}\foo{A}_{\mu}(z, \psi) = \frac{2}{z}\bar{g}_\mu(z, \psi)
	\end{dmath},
\end{dgroup}
where 

\begin{dmath}
		\bar{g}_\mu(z,\psi) \defeq \frac{1}{\pi i}(\mathrm{e}^{iz\sin\psi - i\mu\psi} - 1)
\end{dmath}
fulfils the Nielsen's condition \cref{eq:Nielsen-condition-f=0}. As a consequence, $\foo{A}_{\mu}(z, \psi)$ is a solution of the distinct inhomogeneous Bessel ODE:

\begin{dmath}\label{eq:Anger-Weber-ODE}
	\mathcal{D}_{B_{z,\mu}} \foo{A}_{\mu}(z, \psi) = \bar{\ell}_\mu(z, \psi)
\end{dmath},
specified by the following right-hand-side term (cfr \cref{eq:ell-f=0}):

\begin{dmath}\label{eq:Anger-Weber-ODE-rhs}
	\bar{\ell}_\mu(z, \psi) \defeq \frac{1}{\pi i}\left[ (\mu + z \cos\psi) \mathrm{e}^{i z \sin\psi - i \mu \psi} - z - \mu \right]
\end{dmath}.
More precisely, $\foo{A}_{\mu}(z, \psi)$ is the particular solution satisfying the following initial conditions:

\begin{dgroup}
	\begin{dmath}
		\foo{A}_{\mu}(0, \psi)  = {\frac{1 - \mathrm{e}^{-i \mu \psi}}{i\mu \pi} \equiv -\frac{1}{\mu^2}}\bar{\ell}_\mu(0, \psi)
	\end{dmath},
		\begin{dmath}
		\partial_{z}\foo{A}_{\mu}(0, \psi) = {\frac{(1 - \mathrm{e}^{-i \mu \psi})(\cos\psi + i\mu\sin\psi)}{i\pi(\mu^2 - 1)} \equiv -\frac{1}{\mu^2 - 1}\partial_z\bar{\ell}_\mu(0, \psi)}
	\end{dmath}.
\end{dgroup}

Comparing \cref{eq:Anger-Weber-ODE-rhs} with \cref{eq:Gessel-rhs} and exploiting the linearity of the differential operator $\mathcal{D}_{B_{z,\mu}}$, as well as the form of the distinct inhomogeneous Bessel ODE satisfied by the Anger function, namely:

\begin{dmath}\label{eq:Anger-ODE}
	\mathcal{D}_{B_{z,\mu}} \foo{J}_{\mu}(z) = \frac{\sin\pi\mu}{\pi}(z - \mu)
\end{dmath},
we can express $G_{\mu}(z, \psi)$ in terms of $\foo{J}_{\mu}(z)$ and $\foo{A}_{\mu}(z, \psi)$. It eventually comes out that: 

\begin{dmath}\label{eq:G-JA}
		G_{\mu}(z, \psi)  = \mathrm{e}^{i \mu \psi} \left[\frac{\pi}{\sin\pi\mu}  \foo{J}_{\mu}(z) - i\pi   \foo{A}_{\mu}(-z, \psi) \right] 
\end{dmath}.
Indeed, as expected:
\begin{dmath}
	\mathcal{D}_{B_{z,\mu}} G_\mu(z,\psi) = \mathrm{e}^{i \mu \psi} \left[\frac{\pi}{\sin\pi\mu} \mathcal{D}_{B_{z,\mu}}\foo{J}_{\mu}(z) -i\pi \mathcal{D}_{B_{z,\mu}}\foo{A}_{\mu}(-z,\psi)\right] = 
	\mathrm{e}^{i \mu \psi} \left[z - \mu - (\mu - z\cos\psi) \mathrm{e}^{-i z \sin\psi - i \mu \psi} - z + \mu\right] =
	{(z\cos\psi - \mu) \mathrm{e}^{-i z \sin\psi} \equiv \ell_\mu(z, \psi)}
\end{dmath}.
This result coincides with \cref{eq:G-ODE-simpler} obtained by direct integration in the previous subsection. In virtue of the linearity of the original Cauchy initial value problem, it can be straightforwardly verified that \cref{eq:G-JA} also satisfies the initial conditions  \cref{eq:ic-z} and \cref{eq:ic-psi}.

\noindent We note that \cref{eq:G-JA} obviously entails the following expression for the "rescaled" function $\tilde{G}_{\mu}(z, \psi)$:

\begin{dmath}\label{eq:G-tilde-JA}
		\tilde{G}_{\mu}(z, \psi)  = \foo{J}_{\mu}(z) - i\sin\mu\pi \foo{A}_{\mu}(-z, \psi) 
\end{dmath}.

\section{Application to plasma physics}\label{sec:Vlasov}

The functions $G_\mu(z, \psi)$ arise in plasma kinetic model calculations whenever the method of characteristics is used to find solutions of the Vlasov-Maxwell system of partial differential equations. 
As a concrete application, we present a simplified method for solving the linearised Vlasov PDE that describes a hot magnetised plasma in the presence of a small-amplitude electromagnetic perturbation. The method exploits some of the results reported in the previous sections. 

The linearized Vlasov PDE reads:

\begin{dmath}\label{eq:Vlasov-linear}
	\left[\partial_t+\bv\cdot\partial_{\bx}+\frac{q}{m}(\bv\wedge\bB_0)\cdot\partial_{\bv}\right]f(\bx,\bv,t)=-\frac{q}{m}\left[\bE(\bx,t)+\bv\wedge\bB(\bx,t)\right]\cdot\partial_{\bv}f_0(v_\perp, v_\parallel)
\end{dmath},
where for each species -- species indices are omitted for ease of notation:
	
	\noindent a) $f_{0}(v_\perp, v_\parallel)$ is the unperturbed distribution function, typically a Maxwellian when all the plasma species are initially in thermodynamical equilibrium;
	
	\noindent b) $f(\bm{x},\bm{v},t)$ is the first-order perturbed distribution function;
	
	\noindent c) $\bB_0 = B_0 \hat{\bb}$ is a constant, uniform magnetic field;
	
	\noindent d) $\bE(\bx,t)$ and $\bB(\bx,t)$ describe the small-amplitude electromagnetic perturbation in self-consistent interaction with the plasma.

   The solution of \cref{eq:Vlasov-linear} can be obtained by integrating along characteristics, i.e. unperturbed Larmor orbits. Namely, by preliminarily Fourier-Laplace-transforming\footnote{With a slight abuse of notation we adopt the same symbol for a function and its Fourier-Laplace transform. We also assume that the system occupies a finite volume $V$ with periodic boundary conditions, stipulating to eventually take the limit $V \to \infty$. With these conventions:
	\begin{dmath*}f(\bx, \bv, t) = \frac{1}{V} \sum_\bk{}^' \mathrm{e}^{i\,\bk\cdot\bx}\int_{\mathcal{B}_\alpha} \frac{\mathrm{d}\Omega}{2\pi}f(\bk,\bv,\Omega)\mathrm{e}^{-i\,\Omega_\,t} 
	\end{dmath*} 
	and similarly for $\bE(\bx, t)$ and $\bB(\bx, t)$. Note that $\Omega$ is in general a complex variable. The integral in $\mathrm{d}\Omega$ is carried out along the Bromwich contour $\mathcal{B}_\alpha$ consisting in a line parallel to the real axis and passing trough the point $i\alpha$, where the value of the real parameter $\alpha$ is chosen so as to ensure convergence. The prime in the sum over $\bk$ indicates that the mode $\bk=0$ is excluded.} 
 \cref{eq:Vlasov-linear} and expressing both $\bv$ and $\bk$ in cylindrical coordinates\footnote{$\bv = v_{\perp} \hat{\be}_\perp(\phi) +v_{\parallel} \hat{\bb}$, $\bk = k_{\perp} \hat{\be}_\perp(\phi_\bk) +k_{\parallel} \hat{\bb}$, with $\hat{\be}_\perp(\phi) \defeq \cos\phi\,\hat{\mathbf{e}}_1+\sin\phi\,\hat{\mathbf{e}}_2$ and $\hat{\mathbf{e}}_1$, $\hat{\mathbf{e}}_2$ two arbitrary real versors forming with $\hat{\bb}$ the orthonormal triad $(\hat{\mathbf{e}}_1, \hat{\mathbf{e}}_2, \hat{\bb})$. In order to simplify calculations it is customary to express $\bk$ in the so-called Stix frame, consisting in setting $\phi_\bk=0$, so that the orthogonal component of $\bk$ is oriented along the $x$ axis. This is fully legitimate for the linear equation treated here, where the modes are decoupled, but we prefer to stick to the most general case in view of applications in the nonlinear scenario.}, after some tedious algebra we eventually arrive at the expression:

\begin{dmath}\label{eq:Vlasov-sol}
	f(\bk,\bv,\Omega) = i \frac{q}{m \omega_c}\,\mathrm{e}^{i z_\bk \sin \psi_\bk}c_{\mu_{\bk,\Omega}}\int_0^{2\pi}\frac{\mathrm{d}\lambda}{2\pi}\,\mathrm{e}^{-i z_\bk \sin(\lambda +\psi_\bk) - i \mu_{\bk,\Omega}\lambda}\left\{\mathrm{e}^{i(\lambda + \psi_\bk)}\left[\frac{\mathrm{e}^{i\phi_\bk}}{\sqrt{2}} E_{L}(\bk,\Omega)\,\mathcal{G}_{\bk,\Omega}+\frac{k_\perp}{2\Omega} E_{\parallel}(\bk,\Omega)\mathcal{H}\right] + \mathrm{e}^{-i(\lambda + \psi_\bk)}\left[\frac{\mathrm{e}^{-i\phi_\bk}}{\sqrt{2}} E_{R}(\bk,\Omega)\mathcal{G}_{\bk,\Omega}+\frac{k_\perp}{2\Omega} E_{\parallel}(\bk,\Omega)\mathcal{H}\right] + E_{\parallel}(\bk,\Omega)\partial_{v_\parallel} \right\}f_0(v_\perp, v_\parallel)
\end{dmath}
where $\psi_\bk \defeq \phi -\phi_\bk$, $z_\bk \defeq k_\perp v_\perp/\omega_c$, $\mu_{\bk,\Omega} \defeq (k_\parallel v_\parallel - \Omega)/\omega_c$,

\begin{align}
		E_{R}(\bk,\Omega) &\defeq \frac{E_{1}(\bk,\Omega)+i E_{2}(\bk,\Omega)}{\sqrt{2}}, & E_{L}(\bk,\Omega) &\defeq \frac{E_{1}(\bk,\Omega)-i E_{2}(\bk,\Omega)}{\sqrt{2}}
\end{align}
and we have introduced shorthands for the following differential operators:

\begin{dgroup}
	\begin{dmath}
		\mathcal{G}_{\bk,\Omega} \defeq \partial_{v_\perp} + \frac{k_\parallel}{\Omega}(v_\perp \partial_{v_\parallel} - v_\parallel \partial_{v_\perp})
	\end{dmath},
	\begin{dmath}
		\mathcal{H} \defeq v_\parallel \partial_{v_\perp} - v_\perp \partial_{v_\parallel}
	\end{dmath}.
\end{dgroup}
By comparing \cref{eq:Vlasov-sol} with \cref{eq:G-derivatives} and \cref{eq:G-derivatives+-} we immediately get -- we provisionally drop the indices of $z_\bk$, $\psi_\bk$ and $\mu_{\bk, \Omega}$:

\begin{dmath}\label{eq:Vlasov-sol-G}
	f(\bk,\bv,\Omega) = i \frac{q}{m \omega_c}\mathrm{e}^{i z \sin \psi}\left\{ \\ 
	-\left(\partial_{z} - \frac{i}{z}\partial_{\psi}\right) G_{\mu}(z, \psi)\left[\frac{\mathrm{e}^{i\phi_\bk}}{\sqrt{2}} E_{L}(\bk,\Omega)\mathcal{G}_{\bk,\Omega}+\frac{k_\perp}{2\Omega} E_{\parallel}(\bk,\Omega)\mathcal{H}\right] \\ 
	+ \left(\partial_{z} + \frac{i}{z}\partial_{\psi}\right) G_{\mu}(z, \psi)\left[\frac{\mathrm{e}^{-i\phi_\bk}}{\sqrt{2}} E_{R}(\bk,\Omega)\mathcal{G}_{\bk,\Omega}+\frac{k_\perp}{2\Omega} E_{\parallel}(\bk,\Omega)\mathcal{H}\right] \\
	+ G_{\mu}(z, \psi) E_{\parallel}(\bk,\Omega)\partial_{v_\parallel} \right\}f_0(v_\perp, v_\parallel) =
	i \frac{q}{m \omega_c}\mathrm{e}^{i z \sin\psi}\left\{ \\ 
	\mathrm{e}^{i\psi} G_{\mu - 1}(z, \psi)\left[\frac{\mathrm{e}^{i\phi_\bk}}{\sqrt{2}} E_{L}(\bk,\Omega)\mathcal{G}_{\bk,\Omega}+\frac{k_\perp}{2\Omega} E_{\parallel}(\bk,\Omega)\mathcal{H}\right] \\ 
	+ \mathrm{e}^{-i\psi} G_{\mu + 1}(z, \psi)\left[\frac{\mathrm{e}^{-i\phi_\bk}}{\sqrt{2}} E_{R}(\bk,\Omega)\mathcal{G}_{\bk,\Omega}+\frac{k_\perp}{2\Omega} E_{\parallel}(\bk,\Omega)\mathcal{H}\right] \\
	+ G_{\mu}(z, \psi) E_{\parallel}(\bk,\Omega)\partial_{v_\parallel} \right\}f_0(v_\perp, v_\parallel)
\end{dmath}.

\Cref{eq:Vlasov-sol-G} enables to calculate the plasma equivalent linear susceptibility second-rank tensor $\vec{\chi}(\bk,\Omega)$ in Fourier-Laplace-space. In a a linear, spatially and temporally dispersive, non isotropic medium, the (Fourier-Laplace transform of the) susceptibility tensor satisfies the relation:

\begin{dmath}\label{eq:chi.E}
	\vec{\chi}(\bk, \Omega)\cdot\bE(\bk,\Omega) = \frac{i}{\epsilon_0\Omega}\bj(\bk,\Omega)
\end{dmath}.
The total current density $\bj(\bk,\Omega)$ is defined self-consistently as:

\begin{dmath}\label{eq:j}
	\bj(\bk,\Omega) \defeq \sum_\sigma q_\sigma\int_{-\infty}^{\infty} \bv f_{\sigma}(\bk,\bv,\Omega)\mathrm{d}^3v
\end{dmath},
where we have reintroduced the plasma species index $\sigma$ systematically omitted so far.
Therefore, taking into account the contribution of each plasma species - but dropping again the species index - \cref{eq:chi.E} becomes:

\begin{dmath}\label{eq:chi.E-G}
	\vec{\chi}(\bk, \Omega)\cdot\bE(\bk,\Omega) = 
	-\sum_\sigma\frac{\omega_P^2}{\omega_c\Omega} \int_0^\infty v_\perp\mathrm{d}v_\perp \int_{-\infty}^\infty \mathrm{d}v_\parallel
 		\int_0^{2\pi} \mathrm{d}\phi \,
 		\bv \, \mathrm{e}^{i z \sin\psi}\left\{ \\
	\mathrm{e}^{i\psi} G_{\mu - 1}(z, \psi)\left[\frac{\mathrm{e}^{i\phi_\bk}}{\sqrt{2}} E_{L}(\bk,\Omega)\mathcal{G}_{\bk,\Omega}+\frac{k_\perp}{2\Omega} E_{\parallel}(\bk,\Omega)\mathcal{H}\right]
	+ {\mathrm{e}^{-i\psi} G_{\mu + 1}(z, \psi)\left[\frac{\mathrm{e}^{-i\phi_\bk}}{\sqrt{2}} E_{R}(\bk,\Omega)\mathcal{G}_{\bk,\Omega}+\frac{k_\perp}{2\Omega} E_{\parallel}(\bk,\Omega)\mathcal{H}\right] }
	+ G_{\mu}(z, \psi) E_{\parallel}(\bk,\Omega)\partial_{v_\parallel} \right\}\bar{f}_0(v_\perp, v_\parallel)
\end{dmath}.
For each species, $\omega_P \defeq (q^2 n_0/\epsilon_0 m)^{\frac{1}{2}}$ is the plasma frequency, $n_0$ the constant unperturbed number density and $\bar{f}_0 \equiv f_0/n_0$ the dimensionless, unperturbed distribution function normalized to 1. 

\noindent In \cref{eq:chi.E-G} there appear the "polarized" components of $\bE(\bk, \Omega)$, obtained by decomposing the electric field with respect to the complex orthonormal triad\footnote{
We decompose a generic (usually complex) vector quantity $\bu$ as $\bu = u_1\hat{\be}_1 + u_2\hat{\be}_2 + u_\parallel \hat{\bb} = u_R\hat{\be}_R + u_L\hat{\be}_L + u_\parallel \hat{\bb}$, where $(\hat{\be}_1$, $\hat{\be}_2, \hat{\bb})$ is a real orthonormal triad and
\begin{align*}
	\hat{\be}_R &\defeq \frac{\hat{\be}_1  - i \hat{\be}_2}{\sqrt{2}}, & \hat{\be}_L &\defeq \frac{\hat{\be}_1  + i \hat{\be}_2} {\sqrt{2}}, &
				u_R &\defeq \frac{u_1  + i u_2} {\sqrt{2}}, &
				u_L &\defeq \frac{u_1  - i u_2} {\sqrt{2}}.
\end{align*}
The vectors $(\hat{\be}_R,\,\hat{\be}_L,\,\hat{\bb})$ form an orthonormal complex triad, since $\hat{\be}^*_R\cdot\hat{\be}_R=\hat{\be}^*_L\cdot\hat{\be}_L=1$, $\hat{\be}^*_R\cdot\hat{\be}_L=\hat{\be}^*_L\cdot\hat{\be}_R=\hat{\be}^*_R\cdot\hat{\bb}=\hat{\be}^*_L\cdot\hat{\bb}=0$. We note incidentally that $\hat{\be}^*_R = \hat{\be}_L$ and $\hat{\be}^*_L=\hat{\be}_R$.
}
 $(\hat{\be}_R,\hat{\be}_L,\hat{\bb})$, $\bE(\bk, \Omega) = E_R(\bk, \Omega) \hat{\be}_R + E_L(\bk, \Omega) \hat{\be}_L + E_\parallel(\bk, \Omega)$. 

By expressing also $\bv$ in this basis, the susceptibility tensor in the same basis can be immediately derived by simple inspection. It turns out to consist in the following combination of complex dyadic products:
 
\begin{dmath}\label{eq:chi}
	\vec{\chi}(\bk, \Omega) = -2\pi\sum_\sigma
	\frac{\omega_P^2}{\omega_c\Omega} \int_0^\infty v_\perp\mathrm{d}v_\perp \int_{-\infty}^\infty \mathrm{d}v_\parallel
 		\int_0^{2\pi} \frac{\mathrm{d}\phi}{2\pi} \mathrm{e}^{i z \sin\psi}
 		\bv \bw^*
\end{dmath},
with

\begin{dmath}
	\bv = \frac{v_\perp \mathrm{e}^{i\,\phi}}{\sqrt{2}}\,\hat{\be}_R + \frac{v_\perp \mathrm{e}^{-i\phi}}{\sqrt{2}}\hat{\be}_L + v_\parallel \hat{\bb}
\end{dmath}
and

\begin{dmath}
	\bw^* = w^*_R\hat{\be}^*_R + w^*_L\hat{\be}^*_L + w^*_\parallel \hat{\bb} \equiv 
	G_{\mu + 1}(z, \psi)\frac{\mathrm{e}^{-i\phi}}{\sqrt{2}} \mathcal{G}_{\bk,\Omega}\bar{f}_0 \, \hat{\be}^*_R + 
	G_{\mu - 1}(z, \psi)\frac{\mathrm{e}^{i\phi}}{\sqrt{2}} \mathcal{G}_{\bk,\Omega} \bar{f}_0 \, \hat{\be}^*_L
	+ \left[G_{\mu}(z, \psi) \left(\partial_{v_\parallel} + \frac{\Omega - k_\parallel v_\parallel}{v_\perp \Omega} \mathcal{H} \right)\bar{f}_0 + 
	\frac{\omega_c}{v_\perp \Omega}\mathrm{e}^{-i z \sin\psi}\mathcal{H}\bar{f}_0\right] \, \hat{\bb} =
	G_{\mu + 1}(z, \psi)\frac{\mathrm{e}^{-i\phi}}{\sqrt{2}} \mathcal{G}_{\bk,\Omega}\bar{f}_0 \, \hat{\be}^*_R + 
	G_{\mu - 1}(z, \psi)\frac{\mathrm{e}^{i\phi}}{\sqrt{2}} \mathcal{G}_{\bk,\Omega} \bar{f}_0 \, \hat{\be}^*_L
	+ \left[G_{\mu}(z, \psi) \frac{v_\parallel}{v_\perp} \mathcal{G}_{\bk,\Omega} \bar{f}_0 + 
	\frac{\omega_c}{v_\perp \Omega}\mathrm{e}^{-i z \sin\psi}\mathcal{H}\bar{f}_0\right] \, \hat{\bb}
\end{dmath}.
After expanding the dyadic products and changing the angular integration variable into\footnote{
Note that, since al the integrands are periodic functions of $\psi$ of period $2\pi$, we are legitimated to assume
	\begin{dmath*}
		\int_{0}^{2\pi}\mathrm{d}\phi\,(\dots) = {\int_{-\phi_\bk}^{2\pi-\phi_\bk}\mathrm{d}\psi\,(\dots) \equiv \int_0^{2\pi}\mathrm{d}\psi\,(\dots)}
\end{dmath*}
}
 $\psi = \phi - \phi_\bk$ we obtain:

\begin{dmath}\label{eq:chi-2}
	\vec{\chi}(\bk, \Omega) = -2\pi\sum_\sigma
	\frac{\omega_P^2}{\omega_c\Omega} \int_0^\infty v_\perp\mathrm{d}v_\perp \int_{-\infty}^\infty \mathrm{d}v_\parallel \left\{ \\
 			\frac{v_\perp}{2}\mathcal{G}_{\bk,\Omega}\bar{f}_0\int_0^{2\pi} \frac{\mathrm{d}\psi}{2\pi}\mathrm{e}^{i z \sin\psi} G_{\mu + 1}(z, \psi) \, \hat{\be}_R\hat{\be}^*_R \\
 			+ \frac{v_\perp}{2} \mathrm{e}^{2i\phi_\bk}\mathcal{G}_{\bk,\Omega}\bar{f}_0\int_0^{2\pi} \frac{\mathrm{d}\psi}{2\pi}\mathrm{e}^{i z \sin\psi + 2i \psi} G_{\mu - 1}(z, \psi) \, \hat{\be}_R\hat{\be}^*_L \\
 			+ \frac{v_\parallel}{\sqrt{2}} \mathrm{e}^{i\phi_\bk} \mathcal{G}_{\bk,\Omega}\bar{f}_0\int_0^{2\pi} \frac{\mathrm{d}\psi}{2\pi}\mathrm{e}^{i z \sin\psi + i \psi} G_{\mu}(z, \psi) \, \hat{\be}_R\hat{\bb} +
 			+ \frac{v_\perp}{2} \mathrm{e}^{-2i\phi_\bk} \mathcal{G}_{\bk,\Omega}\bar{f}_0\int_0^{2\pi} \frac{\mathrm{d}\psi}{2\pi}\mathrm{e}^{i z \sin\psi - 2i \psi} G_{\mu + 1}(z, \psi) \, \hat{\be}_L\hat{\be}^*_R \\
 			+ \frac{v_\perp}{2}\mathcal{G}_{\bk,\Omega}\bar{f}_0\int_0^{2\pi} \frac{\mathrm{d}\psi}{2\pi}\mathrm{e}^{i z \sin\psi} G_{\mu - 1}(z, \psi) \, \hat{\be}_L\hat{\be}^*_L \\
 			+ \frac{v_\parallel}{\sqrt{2}} \mathrm{e}^{-i\phi_\bk} \mathcal{G}_{\bk,\Omega}\bar{f}_0\int_0^{2\pi} \frac{\mathrm{d}\psi}{2\pi}\mathrm{e}^{i z \sin\psi - i \psi} G_{\mu}(z, \psi) \, \hat{\be}_L\hat{\bb} \\
 			+ \frac{v_\parallel}{\sqrt{2}} \mathrm{e}^{-i\phi_\bk} \mathcal{G}_{\bk,\Omega}\bar{f}_0\int_0^{2\pi} \frac{\mathrm{d}\psi}{2\pi}\mathrm{e}^{i z \sin\psi - i \psi} G_{\mu + 1}(z, \psi) \, \hat{\bb}\hat{\be}^*_R \\
 			+ \frac{v_\parallel}{\sqrt{2}}\mathrm{e}^{i\phi_\bk} \mathcal{G}_{\bk,\Omega}\bar{f}_0\int_0^{2\pi} \frac{\mathrm{d}\psi}{2\pi}\mathrm{e}^{i z \sin\psi + i \psi} G_{\mu - 1}(z, \psi) \, \hat{\bb}\hat{\be}^*_L \\
 			+ \left( \frac{v_\parallel \omega_c}{v_\perp \Omega} \mathcal{H}\bar{f}_0 + \frac{v^2_\parallel}{v_\perp}\mathcal{G}_{\bk,\Omega}\bar{f}_0\int_0^{2\pi} \frac{\mathrm{d}\psi}{2\pi}\mathrm{e}^{i z \sin\psi} G_{\mu}(z, \psi) \right) \, \hat{\bb}\hat{\bb}
 			\right\}
\end{dmath}.
Expressed in matrix form \cref{eq:chi-2} reads:

\begin{dmath}\label{eq:chi-2-matrix}
	\chi_P(\bk, \Omega) = 2\pi\sum_\sigma
	\frac{\omega_P^2}{\omega_c\Omega} \int_0^\infty v_\perp\mathrm{d}v_\perp \int_{-\infty}^\infty \mathrm{d}v_\parallel\,\mathrm{S}_P(v_\perp,v_\parallel,\bk, \Omega)
\end{dmath},
where  the subscript $P$ identifies matrices in the "polarized" basis and the global minus sign has been absorbed into $\mathrm{S}_P$:

\begin{dmath}\label{eq:S_p}
	\mathrm{S}_P(v_\perp,v_\parallel,\bk, \Omega) = -
		\begin{bmatrix}
    \frac{v_\perp}{2}\mathcal{G}_{\bk,\Omega}\bar{f}_0 I_{1,1} & \frac{v_\perp}{2} \mathrm{e}^{2i\phi_\bk}\mathcal{G}_{\bk,\Omega}\bar{f}_0 I_{1,-1} & \frac{v_\parallel}{\sqrt{2}} \mathrm{e}^{i\phi_\bk} \mathcal{G}_{\bk,\Omega}\bar{f}_0 I_{1,0} \\
    \frac{v_\perp}{2} \mathrm{e}^{-2i\phi_\bk} \mathcal{G}_{\bk,\Omega}\bar{f}_0 I_{-1,1} & \frac{v_\perp}{2}\mathcal{G}_{\bk,\Omega}\bar{f}_0 I_{-1,-1} & \frac{v_\parallel}{\sqrt{2}} \mathrm{e}^{-i\phi_\bk} \mathcal{G}_{\bk,\Omega}\bar{f}_0\ I_{-1,0} \\
    \frac{v_\parallel}{\sqrt{2}} \mathrm{e}^{-i\phi_\bk} \mathcal{G}_{\bk,\Omega}\bar{f}_0 I_{0,1} &  \frac{v_\parallel}{\sqrt{2}}\mathrm{e}^{i\phi_\bk} \mathcal{G}_{\bk,\Omega}\bar{f}_0 I_{0,-1} & \frac{v_\parallel \omega_c}{v_\perp \Omega} \mathcal{H}\bar{f}_0 + \frac{v^2_\parallel}{v_\perp}\mathcal{G}_{\bk,\Omega}\bar{f}_0 I_{0,0}
  \end{bmatrix}
\end{dmath}.
We have adopted the following notation for the integrals entering \cref{eq:S_p}:
\begin{dmath}\label{eq:I_mn}
	I_{m,n} \defeq 
		\int_0^{2\pi} \frac{\mathrm{d}\psi}{2\pi}\mathrm{e}^{i z \sin\psi + i(m - n)\psi} G_{\mu+n}(z, \psi) \equiv
		\frac{\pi}{\sin\mu\pi}\int_0^{2\pi} \frac{\mathrm{d}\psi}{2\pi}\mathrm{e}^{i z \sin\psi + i(\mu + m)\psi} \tilde{G}_{\mu+n}(z, \psi)
\end{dmath}.
The generic integral $I_{m,n}$ is most easily carried out by expressing the function $G_{\mu}(z,\psi)$ in the original form of \cref{eq:G-definite}. It eventually turns out that, for any integer $m$ and $n$:

\begin{dmath}\label{eq:I_mn-sol}
	I_{m,n}  = \frac{\pi}{\sin\mu\pi}
		\begin{cases}
    		(-1)^m J_{\mu + m}J_{-\mu - n} & \text{for } m \geq n \\
   			 (-1)^n J_{\mu + n}J_{-\mu - m} & \text{for } m \leq n
  		\end{cases}
\end{dmath}.
A detailed derivation of \cref{eq:I_mn-sol}, as well as a thorough analysis of its relationships with the Newberger's sum rule, can be found in \cref{sec:integrals-in-chi}. 

\noindent After carrying out the integrals, we obtain for $\mathrm{S}_P$:

\begin{dmath}\label{eq:S_P}
	\mathrm{S}_P(v_\perp,v_\parallel,\bk, \Omega) = \frac{\pi}{\sin\mu\pi}
	\begin{bmatrix}
    	\frac{v_\perp}{2}\mathcal{G}_{\bk,\Omega}\bar{f}_0 J_{\mu + 1} J_{-\mu - 1} & \frac{v_\perp}{2} \mathrm{e}^{2i\phi_\bk}\mathcal{G}_{\bk,\Omega}\bar{f}_0 J_{\mu + 1} J_{-\mu + 1} & \frac{v_\parallel}{\scriptsize{\sqrt{2}}} \mathrm{e}^{i\phi_\bk} \mathcal{G}_{\bk,\Omega}\bar{f}_0 J_{\mu + 1} J_{-\mu} \\
    	\frac{v_\perp}{2} \mathrm{e}^{-2i\phi_\bk} \mathcal{G}_{\bk,\Omega}\bar{f}_0 J_{\mu + 1} J_{-\mu + 1} & \frac{v_\perp}{2}\mathcal{G}_{\bk,\Omega}\bar{f}_0 J_{\mu - 1} J_{-\mu + 1} & -\frac{v_\parallel}{\scriptsize{\sqrt{2}}} \mathrm{e}^{-i\phi_\bk} \mathcal{G}_{\bk,\Omega}\bar{f}_0\ J_{\mu} J_{-\mu + 1} \\
    	\frac{v_\parallel}{\scriptsize{\sqrt{2}}} \mathrm{e}^{-i\phi_\bk} \mathcal{G}_{\bk,\Omega}\bar{f}_0 J_{\mu + 1} J_{-\mu} &  -\frac{v_\parallel}{\scriptsize{\sqrt{2}}}\mathrm{e}^{i\phi_\bk} \mathcal{G}_{\bk,\Omega}\bar{f}_0 J_{\mu} J_{-\mu + 1} & -\frac{\sin\mu\pi}{\pi}\frac{v_\parallel \omega_c}{v_\perp \Omega} \mathcal{H}\bar{f}_0 - \frac{v^2_\parallel}{v_\perp}\mathcal{G}_{\bk,\Omega}\bar{f}_0 J_{\mu} J_{-\mu}
	\end{bmatrix}
\end{dmath}.
Notice that, since $I_{m,n} = I_{n,m}$, only six out of nine integrals are independent. Furthermore, all the products of Bessel functions appearing in \cref{eq:S_P}  can be expressed as combinations of products of only $J_\mu(z)$, $J_{-\mu}(z)$ and their first derivatives with respect to the common argument. We remand to \cref{sec:integrals-in-chi} for details.

The matrix $\chi(\bk, \Omega)$ representing the tensor $\vec{\chi}(\bk, \Omega)$ in standard Cartesian coordinates can be recovered from $\chi_P(\bk, \Omega)$ by means of a simple similarity transformation. Namely,
\begin{dmath}
	\chi(\bk, \Omega) = \mathrm{M}^\dagger\,\chi_P(\bk, \Omega)\,\mathrm{M} = 2\pi\sum_\sigma
	\frac{\omega_P^2}{\omega_c\Omega} \int_0^\infty v_\perp\mathrm{d}v_\perp \int_{-\infty}^\infty \mathrm{d}v_\parallel\,\mathrm{M}^\dagger\,\mathrm{S}_P(v_\perp,v_\parallel,\bk, \Omega)\,\mathrm{M} \equiv 
	2\pi\sum_\sigma
	\frac{\omega_P^2}{\omega_c\Omega} \int_0^\infty v_\perp\mathrm{d}v_\perp \int_{-\infty}^\infty \mathrm{d}v_\parallel\,\mathrm{S}(v_\perp,v_\parallel,\bk, \Omega),
\end{dmath},
where

\begin{dmath}
	\mathrm{M} = 
  \begin{bmatrix}
    \frac{1}{\scriptsize{\sqrt{2}}} &  \frac{i}{\scriptsize{\sqrt{2}}} & 0 \\
    \frac{1}{\scriptsize{\sqrt{2}}} & \frac{-i}{\scriptsize{\sqrt{2}}} & 0 \\
    0 &  0 & 1
  \end{bmatrix}.
\end{dmath}
We eventually obtain the following expressions for the elements of the matrix $\mathrm{S}$\footnote{Note that $\mathrm{S}$ is hermitian when $\Omega$ -- hence $\mu$ -- is real.}:
\begin{dgroup}\label[pluralequation]{eq:S-final-result}
	\begin{dmath}
		S_{11} = v_\perp \mathcal{G}_{\bk,\Omega}\bar{f}_0 \left\{-\frac{\mu^2}{z^2}\frac{\pi}{\sin\mu\pi}J_\mu J_{-\mu} + \frac{\mu}{z^2} + \sin^2\phi_\bk \\\left[\frac{\pi}{\sin\mu\pi}\left(\frac{\mu^2}{z^2}J_\mu J_{-\mu} - J'_\mu J'_{-\mu}\right) - \frac{2\mu}{z^2}\right]\right\}
	\end{dmath}
	\begin{dmath}
		S_{12} = v_\perp \mathcal{G}_{\bk,\Omega}\bar{f}_0 
			\left\{i\,\frac{\mu}{2z}\frac{\pi}{\sin\mu\pi}(J_\mu J_{-\mu})' - \sin\phi_\bk\cos\phi_\bk \\ 
			\left[\frac{\pi}{\sin\mu\pi} \left(\frac{\mu^2}{z^2}J_{\mu} J_{-\mu} - J'_{\mu} J'_{-\mu}\right) -\frac{2\mu}{z^2} \right]\right\}
	\end{dmath}
	\begin{dmath}
		S_{13} = v_\parallel \mathcal{G}_{\bk,\Omega}\bar{f}_0 \left[\cos\phi_\bk \left(\frac{\mu}{z}\frac{\pi}{\sin\mu\pi}J_{\mu} J_{-\mu} - \frac{1}{z}\right) -\frac{i}{2}\sin\phi_\bk \frac{\pi}{\sin\mu\pi}(J_\mu J_{-\mu})'\right]
	\end{dmath}
	\begin{dmath}
		S_{21} = v_\perp \mathcal{G}_{\bk,\Omega}\bar{f}_0 
			\left\{-i\,\frac{\mu}{2z}\frac{\pi}{\sin\mu\pi}(J_\mu J_{-\mu})' - \sin\phi_\bk\cos\phi_\bk \\ 
			\left[\frac{\pi}{\sin\mu\pi} \left(\frac{\mu^2}{z^2}J_{\mu} J_{-\mu} - J'_{\mu} J'_{-\mu}\right) -\frac{2\mu}{z^2} \right]\right\}
	\end{dmath}
	\begin{dmath}
		S_{22} = v_\perp \mathcal{G}_{\bk,\Omega}\bar{f}_0 \left\{-\frac{\pi}{\sin\mu\pi}J'_{\mu} J'_{-\mu} - \frac{\mu}{z^2} - \sin^2\phi_\bk \\\left[\frac{\pi}{\sin\mu\pi}\left(\frac{\mu^2}{z^2}J_\mu J_{-\mu} - J'_\mu J'_{-\mu}\right) - \frac{2\mu}{z^2}\right]\right\}
	\end{dmath}
	\begin{dmath}
		S_{23} = v_\parallel \mathcal{G}_{\bk,\Omega}\bar{f}_0 \left[\frac{i}{2}\cos\phi_\bk \frac{\pi}{\sin\mu\pi}(J_\mu J_{-\mu})' + \sin\phi_\bk \\\left(\frac{\mu}{z}\frac{\pi}{\sin\mu\pi}J_{\mu} J_{-\mu} - \frac{1}{z}\right) \right]
	\end{dmath}
	\begin{dmath}
		S_{31} = v_\parallel \mathcal{G}_{\bk,\Omega}\bar{f}_0 \left[\cos\phi_\bk \left(\frac{\mu}{z}\frac{\pi}{\sin\mu\pi}J_{\mu} J_{-\mu} - \frac{1}{z}\right) +\frac{i}{2}\sin\phi_\bk \frac{\pi}{\sin\mu\pi}(J_\mu J_{-\mu})'\right]
	\end{dmath}
	\begin{dmath}
		S_{32} = v_\parallel \mathcal{G}_{\bk,\Omega}\bar{f}_0 \left[-\frac{i}{2}\cos\phi_\bk \frac{\pi}{\sin\mu\pi}(J_\mu J_{-\mu})' + \sin\phi_\bk \\\left(\frac{\mu}{z}\frac{\pi}{\sin\mu\pi}J_{\mu} J_{-\mu} - \frac{1}{z}\right) \right]
	\end{dmath}
	\begin{dmath}
		S_{33} = -\frac{v_\parallel \omega_c}{v_\perp \Omega} \mathcal{H}\bar{f}_0 - \frac{v^2_\parallel}{v_\perp}\mathcal{G}_{\bk,\Omega}\bar{f}_0 \left(\frac{\pi}{\sin\mu\pi}J_\mu J_{-\mu}\right)
	\end{dmath}
\end{dgroup}

\section{Conclusions}\label{sec:conclusions}
We have shown that the class of functions resulting from the systematisation of the idea initially proposed by Qin \emph{et al.} - which was originally intended as a mere calculation tool - is in fact intimately related to the family of special functions of Bessel type. We have started to explore the rich structure of the novel class of functions, deriving alternative representations and a number of fundamental properties, as well as revealing connections with many branches of the classical theory of special functions. 

As for the application to plasma physics presented in \cref{sec:Vlasov}, we remark that \Cref{eq:S-final-result} generalize the results reported in Qin \emph{et al.} \cite{hQ07}, which are limited to the special case $\phi_\bk=0$ corresponding to the adoption of Stix coordinates\footnote{We signal the presence of a typo in eq.~35 of Qin \emph{et al.} \cite{hQ07}: elements $(1,2)$ and $(2,1)$ of the tensor $\bT$ should be interchanged}. The same results can be obtained\cite{dgS03}, yet with a considerably greater algebraic effort, by following the usual approach based on the Jacobi-Anger identity \cref{eq:Jacobi-Anger} and then eliminating all the resultant infinite sums by using the Newberger's sum rule (see \cref{sec:integrals-2}), which enables to express series of products of two Bessel functions of integer order in terms of products of two Bessel functions of non-integer (possibly complex) order.

\noindent Apart from the unavoidable algebraic complexity of the resulting formulas, the new procedure outlined here greatly simplifies calculations and helps shed light on the rich mathematical structure underlying the problem. Plans are ongoing to extend this approach to the much more complex nonlinear domain.

\begin{acknowledgments}
We wish to acknowledge the support of Dr. G. Dattoli in offering suggestions and encouragement.
\end{acknowledgments}

\appendix

\section{Derivation of \cref{eq:S-Anger}}\label{sec:S-Anger}
We introduce the shorthand notation $\bar{G}_\mu(z) \defeq G_\mu(z,0)$. Starting from the recurrence relations for cylindrical Bessel functions:

\begin{dgroup}
	\begin{dmath}
		2\partial_z J_n(z) = J_{n-1}(z)-J_{n+1}(z)
	\end{dmath}
		\begin{dmath}
		\frac{2n}{z} J_n(z) = J_{n-1}(z)+J_{n+1}(z)
	\end{dmath},
\end{dgroup}
if we multiply by $\frac{1}{n + \mu}$, sum over $n$ and multiply the resulting equations by $z$, we get:

\begin{dgroup}\label[pluralequation]{eq:S-recurrence-rels}
	\begin{dmath}\label{eq:S-recurrence-diff}
		2z\partial_z \bar{G}_{\mu}(z) = z\bar{G}_{\mu + 1}(z) - z\bar{G}_{\mu - 1}(z)
	\end{dmath}
		\begin{dmath}\label{eq:S-recurrence-sum}
		-2\mu\bar{G}_{\mu}(z) = z\bar{G}_{\mu + 1}(z) + z\bar{G}_{\mu - 1}(z) - 2
	\end{dmath}.
\end{dgroup}
By summing and subtracting \cref{eq:S-recurrence-rels} we obtain:

\begin{dgroup}
	\begin{dmath}\label{eq:S-recurrence-sum-2}
		\left(z\partial_z - \mu\right)\bar{G}_{\mu}(z) = z\bar{G}_{\mu + 1}(z) - 1
	\end{dmath}
		\begin{dmath}\label{eq:S-recurrence-diff-2}
		\left(z\partial_z + \mu\right)\bar{G}_{\mu}(z) = -z\bar{G}_{\mu - 1}(z) + 1
	\end{dmath}.
\end{dgroup}
If we apply $\left(z\frac{d}{dz} + \mu\right)$ to \cref{eq:S-recurrence-sum-2} and use \cref{eq:S-recurrence-diff-2} to eliminate $\bar{G}_{\mu + 1}(z)$, we eventually get a differential equation for $\bar{G}_{\mu}(z)$, namely:

\begin{dmath}\label{eq:Anger-diff-eq}
	\left(z^2\partial_z^2 + z\partial_z +z^2 - \mu^2\right)\bar{G}_{\mu}(z) = z - \mu
\end{dmath}.

This is an inhomogeneous ODE of Bessel type, known as Anger differential equation. The solution of \cref{eq:Anger-diff-eq}  with initial conditions

\begin{dgroup}
	\begin{dmath}
		\bar{G}_{\mu}(0) =  {\sum_n \frac{J_n(0)}{n+\mu} = 
		\sum_n \frac{\delta_{n,0}}{n+\mu} =
		\frac{1}{\mu}}
	\end{dmath}
		\begin{dmath}
		\frac{d}{dz}\bar{G}_{\mu}(0) = {\frac{\bar{G}_{\mu + 1}(0) - \bar{G}_{\mu - 1}(0)}{2} = 
		\frac{1}{1 - \mu^2}}
	\end{dmath}.
\end{dgroup}
is \cite{nW66}:

\begin{dmath}
	\bar{G}_{\mu}(z) = \frac{\pi}{\sin\pi \mu}\foo{J}_{\mu}(z) 
\end{dmath}.

The validity of this result can also be verified directly from the definition given in \cref{eq:G-definite}. Namely:

\begin{dmath}
	\frac{\sin\pi \mu}{\pi} G_\mu(z, 0) =  
	\int_0^{2\pi}\frac{\mathrm{d}\lambda}{2\pi}\,\mathrm{e}^{-i z \sin\lambda - i  \mu (\lambda - \pi)} =
	\int_{-\pi}^{\pi}\frac{\mathrm{d}\lambda}{2\pi}\,\mathrm{e}^{i z \sin\lambda - i  \mu \lambda} =
	\frac{1}{2}\left(
		\int_{0}^{\pi}\frac{\mathrm{d}\lambda}{\pi}\,\mathrm{e}^{i z \sin\lambda - i  \mu \lambda} +  
		\int_{-\pi}^{0}\frac{\mathrm{d}\lambda}{\pi}\,\mathrm{e}^{i z \sin\lambda - i  \mu \lambda}\right) =
	\frac{1}{2}\left(\foo{J}_{\mu}(z) - i\foo{E}_{\mu}(z) + \int_{0}^{\pi}\frac{\mathrm{d}\lambda}{\pi}\,\mathrm{e}^{i z \sin(\lambda - \pi) - i  \mu (\lambda - \pi)}\right) = 
	\frac{1}{2}\left(\foo{J}_{\mu}(z) - i\foo{E}_{\mu}(z) + 
		\mathrm{e}^{i\pi \mu}\int_{0}^{\pi}\frac{\mathrm{d}\lambda}{\pi}\,\mathrm{e}^{-i z \sin\lambda - i  \mu\lambda}\right) = 
	\frac{1}{2}\left[\foo{J}_{\mu}(z) - i\foo{E}_{\mu}(z) + \mathrm{e}^{i\pi \mu}\left(\foo{J}_{-\mu}(z) + i\foo{E}_{-\mu}(z)\right)\right] =
	\frac{1}{2}\left[\foo{J}_{\mu}(z) + \cos\pi \mu \foo{J}_{-\mu}(z) - \sin\pi \mu\foo{E}_{-\mu}(z) - i\left(\foo{E}_{\mu}(z) - \sin\pi \mu\foo{J}_{-\mu}(z) -\cos\pi \mu\foo{E}_{-\mu}(z)\right)\right] =
	\foo{J}_{\mu}(z)
\end{dmath}.
This obviously also entails the useful identity:
\begin{dmath}
	\foo{J}_{\mu}(z) \defeq \mathrm{e}^{i\pi\mu}\int_{0}^{2\pi} \frac{\mathrm{d}\lambda}{2\pi}\,\mathrm{e}^{-i z \sin\lambda - i  \mu \lambda}
\end{dmath}
In the verification use has been made of the definition of the Weber function
\begin{dmath}
	\foo{E}_{\mu}(z) \defeq \int_0^{\pi} \frac{\mathrm{d}\lambda}{\pi}  \sin(\mu\lambda - z \sin\lambda)
\end{dmath},
as well as of the following identities \cite{nW66}: 

\begin{dgroup}
	\begin{dmath}
		\foo{J}_{\mu}(z) = \cos\pi \mu\foo{J}_{-\mu}(z) - \sin\pi \mu\foo{E}_{-\mu}(z)
	\end{dmath}
	\begin{dmath}
		\foo{E}_{\mu}(z) = \cos\pi \mu\foo{E}_{-\mu}(z) + \sin\pi \mu\foo{J}_{-\mu}(z)
	\end{dmath}.
	\end{dgroup}

\section{Calculation of the integrals $I_{m,n}$ entering \cref{eq:chi-2}}\label{sec:integrals-in-chi}

\subsection{Integral representation for the product of two Bessel functions}\label{sec:integrals-1}

As a preliminary step, we recall that the product of two Bessel function of identical argument and different, not necessarily integer indices $\bar{\mu}$ and $\bar{\nu}$ admits the following integral representation \cite[Eq.~10.9.26]{dlmf}, valid for $\mathrm{Re}\,(\bar{\mu} + \bar{\nu}) > -1$:

\begin{dmath}\label{eq:cauchy}
	J_{\bar{\mu}}(z) J_{\bar{\nu}}(z) = 2\int_0^{\frac{\pi}{2}} \frac{\mathrm{d}\theta}{\pi}\,J_{\bar{\mu}+\bar{\nu}}(2z \cos\theta) \cos(\bar{\mu}-\bar{\nu})\theta
\end{dmath}.
In particular, when $\bar{\mu} = \mu + m$, $\bar{\nu} = -\mu - n$, with $m$ and $n$ any integer numbers such that $\mathrm{Re}\,(\bar{\mu} + \bar{\nu}) = m-n \geq 0$, \cref{eq:cauchy} becomes:

\begin{dmath}
	J_{\mu + m}(z) J_{-\mu - n}(z) = 
		2\int_0^{\frac{\pi}{2}} \frac{\mathrm{d}\theta}{\pi}\,J_{m-n}(2z \cos\theta) \cos2\left(\mu + \frac{m+n}{2}\right)\theta =
		\int_{-\pi}^{\pi}\frac{\mathrm{d}\theta}{2\pi}\,J_{m-n}\left(2z \cos\frac{\theta}{2}\right) \mathrm{e}^{-i\left(\mu + \frac{m+n}{2}\right)\theta} =
		\mathrm{e}^{i\left(\mu + \frac{m+n}{2}\right)\pi}\int_{0}^{2\pi} \frac{\mathrm{d}\theta}{2\pi}\,J_{m-n}\left(2z \sin\frac{\theta}{2}\right) \mathrm{e}^{-i\left(\mu + \frac{m+n}{2}\right)\theta}
\end{dmath}.
When $m-n \leq 0$, we simply exchange $m$ with $n$, obtaining:

\begin{dmath}
	J_{\mu + n}(z) J_{-\mu - m}(z) = 
		\mathrm{e}^{i\left(\mu + \frac{m+n}{2}\right)\pi}\int_{0}^{2\pi} \frac{\mathrm{d}\theta}{2\pi}\,J_{n-m}\left(2z \sin\frac{\theta}{2}\right) \mathrm{e}^{-i\left(\mu + \frac{m+n}{2}\right)\theta} =
		(-1)^{m-n}\mathrm{e}^{i\left(\mu + \frac{m+n}{2}\right)\pi}\int_{0}^{2\pi} \frac{\mathrm{d}\theta}{2\pi}\,J_{m-n}\left(2z \sin\frac{\theta}{2}\right) \mathrm{e}^{-i\left(\mu + \frac{m+n}{2}\right)\theta}
\end{dmath},
where in the last passage use has been made of the well-known identity $J_{-n}(z) = (-1)^n J_n(z)$, which holds for any integer $n$.

\noindent In summary, we have established the following intermediate result:

\begin{dmath}\label{eq:intermediate}
	\mathrm{e}^{i\left(\mu + \frac{m+n}{2}\right)\pi}\int_{0}^{2\pi} \frac{\mathrm{d}\theta}{2\pi}\,J_{m-n}\left(2z \sin\frac{\theta}{2}\right) \mathrm{e}^{-i\left(\mu + \frac{m+n}{2}\right)\theta} = 
		\begin{cases}
 			J_{\mu + m}J_{-\mu - n} & \text{for } m \geq n \\
			(-1)^{n-m} J_{\mu + n}J_{-\mu - m} & \text{for } m \leq n
  		\end{cases}
  \end{dmath}.

\subsection{Calculation of $I_{m,n}$}\label{sec:integrals-2}

Let us now turn to the calculation of the integrals $I_{m,n}$. Using simple trigonometric rules and taking into account the definition of $G_\mu(z,\psi)$ \cref{eq:G-definite} and the periodicity of $c_\mu$ with unit period, we can write:

\begin{dmath}
	I_{m,n} = \int_0^{2\pi} \frac{\mathrm{d}\psi}{2\pi}\mathrm{e}^{i z \sin\psi + i(m - n)\psi} G_{\mu+n}(z, \psi) =
	c_{\mu + n}\int_0^{2\pi} \frac{\mathrm{d}\lambda}{2\pi} \mathrm{e}^{-i (\mu + n)\lambda} \int_0^{2\pi} \frac{\mathrm{d}\psi}{2\pi}\mathrm{e}^{-i z[\sin(\lambda + \psi) -\sin\psi] + i(m - n)\psi} =
	c_{\mu}\int_0^{2\pi} \frac{\mathrm{d}\lambda}{2\pi} \mathrm{e}^{-i (\mu + n)\lambda} \int_0^{2\pi} \frac{\mathrm{d}\psi}{2\pi}\mathrm{e}^{-i \left(2z\sin\frac{\lambda}{2}\right)\cos\left(\psi + \frac{\lambda}{2}\right) + i(m - n)\psi}
\end{dmath}
If we write $c_\mu$ explicitly (cf. \cref{eq:c_mu}) and perform the change of variable $\psi \mapsto \theta \defeq \psi +  \frac{\lambda}{2} + \frac{\pi}{2}$, we obtain:

\begin{dmath}\label{eq:lastline}
	I_{m,n} = 
	\frac{\pi}{\sin\mu\pi} \mathrm{e}^{i\left(\mu - \frac{m-n}{2}\right)\pi}\int_0^{2\pi} \frac{\mathrm{d}\lambda}{2\pi} \mathrm{e}^{-i \left(\mu + \frac{m+n}{2}\right)\lambda} \int_0^{2\pi} \frac{\mathrm{d}\theta}{2\pi}\mathrm{e}^{-i \left(2z\sin\frac{\lambda}{2}\right)\sin\theta + i(m - n)\theta} =
	(-1)^m \frac{\pi}{\sin\mu\pi} \left[\mathrm{e}^{i\left(\mu + \frac{m+n}{2}\right)\pi}\int_0^{2\pi} \frac{\mathrm{d}\lambda}{2\pi} J_{m-n}\left(2z\sin\frac{\lambda}{2}\right)\mathrm{e}^{-i \left(\mu + \frac{m+n}{2}\right)\lambda}\right]
\end{dmath},
where in the last passage we have used the following integral representation for Bessel functions of integer order:

\begin{dmath}
	J_n(x) = \int_0^{2\pi} \frac{\mathrm{d}\theta}{2\pi}\mathrm{e}^{-i x\sin\theta + i n\theta}
\end{dmath}.
It is immediate to recognize that the term in square brackets in the last line of \cref{eq:lastline} is identical to the the left hand side of \cref{eq:intermediate}. This enables, in conclusion, to establish the final result:

\begin{dmath}
	I_{m,n} = \frac{\pi}{\sin\mu\pi} 
  		\begin{cases}
    		(-1)^{m} J_{\mu + m}J_{-\mu - n} & \text{for } m \geq n \\
			(-1)^{n} J_{\mu + n}J_{-\mu - m} & \text{for } m \leq n
  		\end{cases}
\end{dmath}.

\subsection{Connection with the Newberger's sum rule}\label{sec:newberger}

The integrals $I_{m,n}$ can also be carried out by exploiting the Jacobi-Anger identity \cref{eq:Jacobi-Anger}. Namely, since

\begin{dmath}
		\tilde{G}_\mu(z, \psi) = \frac{\sin\mu\pi}{\pi}\sum_l \frac{J_l(z)}{l + \mu}\mathrm{e}^{-i(\mu+l)\psi}
	\end{dmath}
we have:

\begin{dmath}\label{eq:I_mn-Newberger}
	I_{m,n} =  
		\frac{\pi}{\sin\mu\pi}\int_0^{2\pi} \frac{\mathrm{d}\psi}{2\pi}\mathrm{e}^{i z \sin\psi + i(\mu + m)\psi} \tilde{G}_{\mu+n}(z, \psi) = 
		\sum_l \frac{J_l(z)}{l + \mu + n}\int_0^{2\pi} \frac{\mathrm{d}\psi}{2\pi}\mathrm{e}^{i z \sin\psi + i(m - n - l)\psi} =
		\sum_l \frac{J_l(z) J_{l + n - m}\,(z)}{l + n + \mu}
\end{dmath}.
The last line of the previous equation is a particular case of the so-called Turkin's function:

\begin{dmath}
	T_M(z,\alpha) \defeq \sum_l \frac{J_{l}(z) J_{l - M}\,(z)}{l + \alpha}
\end{dmath}
corresponding to the choice $M = m - n$, $\alpha = \mu +n$. Newberger \cite{bsN82} and Bakker and Temme \cite{nB84} demonstrated the following identity, known as Newberger's sum rule:

\begin{dmath}\label{eq:newberger}
	T_M(z,\alpha) = \frac{\pi}{\sin\alpha\pi}
	\begin{cases}
		(-1)^M J_{\alpha + M}(z)J_{-\alpha}(z) \condition*{M\ge 0} \\
		J_{-\alpha - M}(z)J_{\alpha}(z) \condition*{M\le 0}
	\end{cases}
\end{dmath}.
It is immediate to verify that, once applied to \cref{eq:I_mn-Newberger}, this identity enables to reproduce the result \cref{eq:I_mn-sol}, derived by other means in this article.

On the other hand, if we start from \cref{eq:I_mn-sol}, \cref{eq:I_mn-Newberger} can also be considered as a generalization of \cref{eq:newberger}. Namely, since:

\begin{dmath}
	I_{m,n} = 
  		\sum_l \frac{J_l(z) J_{l + n - m}\,(z)}{l + n + \mu} =
		\sum_r \frac{J_{r - n}(z) J_{r - m}\,(z)}{r + \mu}
\end{dmath},
the following generalized Newberger's sum rule holds:

\begin{dmath}
	\sum_r \frac{J_{r - n}(z) J_{r - m}\,(z)}{r + \mu} = 
		\frac{\pi}{\sin\mu\pi}
  			\begin{cases}
    			(-1)^{m} J_{\mu + m}J_{-\mu - n} & \text{for } m \geq n \\
				(-1)^{n} J_{\mu + n}J_{-\mu - m} & \text{for } m \leq n
  			\end{cases}
\end{dmath}.

\subsubsection{Expressions in terms of $J_\mu(z)$, $J_{-\mu}(z)$ and their first derivatives}\label{sec:integrals-3}

We notice that, by using the identities:

\begin{dgroup}
	\begin{dmath}
		J_{\mu - 1}(z) = \frac{\mu}{z}J_\mu(z) + J'_\mu(z)
	\end{dmath},
	\begin{dmath}
		J_{\mu + 1}(z) = \frac{\mu}{z}J_\mu(z) - J'_\mu(z) 
	\end{dmath},
\end{dgroup}
 all the integrals entering \cref{eq:chi-2} can be eventually written in terms of only $J_\mu(z)$, $J_{-\mu}(z)$ and their first derivatives with respect to $z$. 
 
 \noindent In the following we report for reference the relevant formulas (omitting for sake of conciseness the common argument of the Bessel functions):

\begin{dmath}
	I_{1,1} = 
		{-\frac{\pi}{\sin\mu\pi} J_{\mu + 1} J_{-\mu - 1} = 
		\frac{\pi}{\sin\mu\pi}\left(\frac{\mu^2}{z^2}J_{\mu} J_{-\mu} - \frac{\mu}{z}(J_{\mu} J_{-\mu})' + J'_{\mu} J'_{-\mu}\right)}
\end{dmath}

\begin{dmath}
	I_{1,-1} = 
		{-\frac{\pi}{\sin\mu\pi}J_{\mu + 1} J_{-\mu + 1} = 
		-\frac{\pi}{\sin\mu\pi} \left(\frac{\mu^2}{z^2}J_{\mu} J_{-\mu} + J'_{\mu} J'_{-\mu}\right) -\frac{2\mu}{z^2}}
\end{dmath}

\begin{dmath}
	I_{1,0} = 
		{-\frac{\pi}{\sin\mu\pi} J_{\mu + 1} J_{-\mu} = 
		\frac{\pi}{\sin\mu\pi} \left(\frac{\mu}{z}J_{\mu}J_{-\mu} -J'_\mu J_{-\mu}\right)}
\end{dmath}

\begin{dmath}
	I_{-1,1} = I_{1,-1} 
\end{dmath}

\begin{dmath}
	I_{-1,-1} = 
	{-\frac{\pi}{\sin\mu\pi} J_{\mu - 1} J_{-\mu + 1} = 
	\frac{\pi}{\sin\mu\pi} \left(\frac{\mu^2}{z^2}J_{\mu} J_{-\mu} + \frac{\mu}{z}(J_{\mu} J_{-\mu})' + J'_{\mu} J'_{-\mu}\right)}
\end{dmath}

\begin{dmath}
	I_{-1,0} = 
		{\frac{\pi}{\sin\mu\pi}J_{\mu} J_{-\mu + 1} =
		\frac{\pi}{\sin\mu\pi}\left(-\frac{\mu}{z}J_{\mu}J_{-\mu} - J_{\mu}J'_{-\mu}\right)}
\end{dmath}

\begin{dmath}
	I_{0,1} = I_{1,0}
\end{dmath}

\begin{dmath}
	I_{0,-1} = I_{-1,0}
\end{dmath}

\begin{dmath}
	I_{0,0} = 
		\frac{\pi}{\sin\mu\pi} J_{\mu} J_{-\mu}
\end{dmath}

\nocite{*}
\bibliography{mybib}

\end{document}